# Illustrating a neural model of logic computations: The case of Sherlock Holmes' old maxim*

Eduardo Mizraji

Group of Cognitive Systems Modeling, Biophysics Section, Facultad de Ciencias,
Universidad de la República, Iguá 4225, Montevideo 11400, Uruguay
E-mails: mizraj@fcien.edu.uy, emizraji@gmail.com

ABSTRACT: Natural languages can express some logical propositions that humans are able to understand. We illustrate this fact with a famous text that Conan Doyle attributed to Holmes: *"It is an old maxim of mine that when you have excluded the impossible, whatever remains, however improbable, must be the truth"*. This is a subtle logical statement usually felt as an evident truth. The problem we are trying to solve is the cognitive reason for such a feeling. We postulate here that we accept Holmes' maxim as true because our adult brains are equipped with neural modules that naturally perform modal logical computations.

**Keywords**: *Neural computations; Natural language; Models of reasoning; Modal logics*

* Previous version in arXiv:
 "The neural computation of Sherlock Holmes' old maxim"

# 1. Introduction

Human language is used to express logical and mathematical computations whose cognitive bases still remain unexplained. We ask ourselves if language acquisition involves a kind of implicit logical and mathematical programming that could explain such performances. Examples of these performances are some logical propositions, transferred by natural language, valid for different languages and for large populations of humans sharing similar cultural traditions. In the present work we choose –as a largely accepted logical statement– one of the most cited expressions that Arthur Conan Doyle attributed to Sherlock Holmes, the "old maxim" mentioned by the character in the story "The Adventure of the Beryl Coronet". For an allusion to this maxim in a scientific context see Cairns-Smith (1990). We enunciate different versions of this "old maxim", and the reader can explore by himself the origin of the conviction that these expressions usually provoke (see Doyle 1988). This maxim is a subtle logical statement and the interesting (and even astonishing) point is that many readers feel that this maxim is an evident truth. One problem we are trying to explain is the following: which is the cognitive reason for such a feeling? The method we follow here implies to keep the theories as simple as possible. We avoid in this first approach the temptation to deeply expand Bayesian theories, logical formalisms or the mathematical technicalities of neural models. Our aim is to explore the possibility of establishing an acceptable link between very different disciplines that are, however, all connected with our problem.

Perhaps the most important question about the "old maxim" concerns the consistency of the statement. At the same time, the most enigmatic aspect of this maxim is the natural acceptability triggered by it in our mind. We are going to analyze two approaches to consistency: the logical consistency and the "neurocomputational consistency". We propose to accept the logical consistency whenever the statement can be immersed in normal well-formed formulas of basic propositional and modal logic. In addition, we propose to accept the neurocomputational acceptability if the logical computation of the statement can be executed by a neural network model. In this work we intend to show that both consistency conditions are satisfied. We begin by describing a Bayesian approach to the maxim. Then, we provide a logical framework to Holmes' statement. After this, we show how these logical approaches can be computed in model neural modules capable to integrate modular networks (ie. networks of networks) that execute a variety of logical operations. With these purposes, we begin describing some basic concepts of propositional calculus and modal logic that confer the logical support of Holmes' maxim. Then, we present a kind of modular neural model based on matrix algebra, and describe how logic operations can be very naturally incorporated into these matrix modules. Finally, we put together this material to provide a neural model that is capable to represent Holmes' maxim. This approach allows us to propose an explanation of the reason why the Holmes' maxim seems so naturally acceptable by us. Our explanation is based on modeling some particular neural networks capable of representing the cognitive computation of modalities.

## 2. Holmes' Old Maxim

Perhaps the most cited version is the following: *"It is an old maxim of mine that when you have excluded the impossible, whatever remains, however improbable, must be the truth"* (in Appendix 1 we present three versions of this maxim with the corresponding sources). In the articles compiled by Eco and Sebeok (1988) there are many allusions to these texts, as well as some formalizations of the methods attributed by Conan Doyle to Holmes using techniques coming from mathematical logic.

The involvement of probability in this text invites us to approach the meaning of these statements using a naïve Bayesian description. In this context, the probability $P(A_i | B)$ of $A_i$ being the cause of event B is given by

$$P(A_i | B) = \frac{P(B | A_i) P(A_i)}{\sum_{j=1}^{n} P(B | A_j) P(A_j)} \qquad (1)$$

where it is assumed that the set $\{A_1, \ldots, A_n\}$ includes disjoint events susceptible of being interpreted as causes of the event B (eg. B can be a symptom of a disease and $\{A_1, \ldots, A_n\}$ the set of potential pathological agents). Each $P(B | A_j)$ measures the conditional probability of event B given $A_j$, and each $P(A_j)$ measures (or estimates) the *a priori* probability of event $A_j$.

In this Bayesian framework, Holmes' maxim can be interpreted as follows: after a research process the investigator can establish that for all $j \neq i$ $P(B | A_j) \to 0$: As a consequence, only $P(B | A_i) \neq 0$. Hence, at this stage, Bayes' formula adopts the following aspect:

$$P(A_i | B) \approx \frac{P(B | A_i) P(A_i)}{0 + \cdots + P(B | A_i) P(A_i) + \cdots + 0} \qquad (2)$$

Consequently, $P(A_i | B) \to 1$ independently of the value of the *a priori* probability $P(A_i)$. This deliberately simplified Bayesian approach gives us a good insight into the meaning of Holmes' maxim.

It is important to remark that, together with allusions to probability, Holmes' maxim includes the words "impossible" and "truth". These two words point to a logical framework, and particularly "impossible" leads us to the domain of modal logic. Consequently, we are going to assume that understanding our cognitive acceptation of the maxim requires to start from logic, and to include *a posteriori* a probabilistic argument inside the logical frame. Moreover, we need to include this logical frame inside a neural model in order to explore how a biological device such as the brain could become prone to accept the validity of Holmes' maxim. We introduce the neural modeling framework in Section 4.1 and the probabilistic approach returns in Section 4.3.

## 3. Logical Consistency

Some of the fundamental ideas about modal logic were exposed by Aristotle in his short text "On Interpretation" (Aristotle, around 350 BC, edition E.M. Edghill). The two basic modal operators "possibility" and "necessity" are formally related by a postulate clearly stated by Aristotle. We provide here with a version of this postulate (Aristotle, 350 BC, Chapter 13):

*"It remains, therefore, that the proposition 'it is not necessary that it should not be' follows from the proposition 'it may be' "*.

A modern formalization of modal logic can be found in Hugues and Cresswell 1972. The symbolic representation for the previous postulate is the following,

$$\Diamond(Q) \equiv \neg\Box\left[\neg(Q)\right] \quad\quad\quad (3)$$

where $\Diamond$ represents the modal proposition "It is possible", $\Box$ means "It is necessary", $\neg$ is the negation, and $\equiv$ represents the logical equivalence. Q represents any proposition.

An equivalent representation is

$$\Box(Q) \equiv \neg\Diamond\left[\neg(Q)\right] \quad\quad\quad (4)$$

A corollary, based on the fact that the double negation corresponds to an affirmation, is

$$\neg[\Diamond(Q)] \equiv \Box[\neg(Q)] \qquad (5)$$

Note that if Q is an arbitrary proposition, its negation $\neg(Q)$ can be interpreted as "whatever remains" once excluded Q. Hence, this corollary (5) shows the proximity of Aristotle's postulate with Holmes' maxim: If it is true that Q is impossible, it is true that what remains, $\neg(Q)$, is necessary.

Remark that the equivalences (3), (4) and (5) are true in all cases, notwithstanding the truth-value of the modal evaluations of Q (i.e., both members of the equivalence can be false). The relation with Holmes' maxim restricts to the case in which both members of the equivalence are true.

The works of De Morgan, Gregory, Boole and Peirce, among others, were fundamental to transform logic into a discipline susceptible of being supported by the techniques of mathematics. Their work produced a variety of mathematical representations of logical calculus. We note here that, deeply influenced by the symbolic methods developed in the field of differential equations (his main area of expertise), Boole established the basic conditions that allow to map the logic truth-values on mathematical variables and to transform the logical statements in mathematical functions (Boole 1847, 1854). In addition, in his book "The Laws of Thought" (1854) Boole attempted to link logical procedures with probability theory. From this Boolean approach, it was possible to define functional relations for all the fundamental logical operations (e.g., negation, disjunction, conjunction, implication, exclusive-or, equivalence).

When logic is immersed into this "Boole's Universe" (BU), the truth-values define a set

$$\tau_2 = \{\, t, f \,\}$$

where t and f are abstract (even arbitrary) mathematical objects corresponding to the truth-values "true" and "false" respectively. In this BU, the monadic functions Mon are applications

$$\text{Mon} : \tau_2 \to \tau_2$$

(the negation Not is an example of these monadic functions, being Not (t) = f and Not (f) = t). The dyadic functions Dyad are applications

$$\text{Dyad} : \tau_2 \times \tau_2 \to \tau_2 \;;$$

where the symbol × indicates the Cartesian product. The truth tables used to represent the logical operations (e.g. disjunction or implication) are examples of the use of these dyadic mappings.

Following the methods created by Boole, other researchers tried to represent modal operations as mathematical functions. Nevertheless, in a famous work Łukasiewicz (1930) demonstrated the impossibility to represent "possibility" and "necessity" as mathematical functions inside the two-valued logic defined in BU (for details, see Łukasiewicz 1930, Mizraji 2008). Consequently, the search for truth-functional representations for these logical modalities leads Łukasiewicz to extend the truth-value space by adding a third value "u" corresponding to uncertain or undecidable propositions. In this new Łukasiewicz's Universe (LU), the logical monadic and dyadic functions are built up over the set

$$\tau_3 = \{\, t,\, f,\, u \,\}$$

where

$$\text{Mon} : \tau_3 \to \tau_3$$

and

$$\text{Dyad} : \tau_3 \times \tau_3 \to \tau_3 \ .$$

Inside this LU, the classical modalities become monadic logical functions, and "possibility" and "necessity" can be respectively expressed by the functions possibility $\Diamond(x)$ and necessity $\Box(x), x \in \tau_3$, defined as follows:

$$\Diamond(t) \equiv \Diamond(u) \equiv t \ ; \ \Diamond(f) = f$$
$$\Box(t) \equiv t \ ; \ \Box(u) \equiv \Box(f) \equiv f \quad ,$$

The negation is defined in the LU as follows:

$$\neg(t) \equiv f \ ; \ \neg(u) \equiv u \ ; \ \neg(f) \equiv t \quad .$$

With this formal repertoire, we can represent Aristotle's postulate with a truth-functional equivalence:

$$\Diamond[x(Q)] \equiv \neg \Box \left[ \neg (x(Q)) \right] \qquad (6)$$

where $x(Q) \in \tau_3$ is the truth-value corresponding to an abstract proposition Q. Equation (6) transforms the Aristotelian postulate into a theorem. Being the negation $\neg$ an idempotent operator $\left(\neg[\neg(x)] \equiv x\right)$, we can deduce the following equivalence from (6):

$$\neg\left[\Diamond[x(Q)]\right] \equiv \Box\left[\neg(x(Q))\right] \qquad (7)$$

We are going to assume the following Axiom:

AXIOM: $\neg(x(Q)) \equiv x(\text{Negation } Q)$ ;

Observe that, in this truth-functional format, our logical operators apply only to truth-values; instead "Negation Q" designates the negation of a proposition. For instance, the truth-value of Q = "*3 is an even number*" is f and its negation is t; the truth-value of Negation Q = "*3 is not an even number*" is t; "Negation Q" is not a mathematical variable but a proposition and can be considered as a propositional definition of negation. This Axiom can be proved as a Lemma (Mizraji and Lin 2011) if we consider that Q refers to a category W (a set) of propositions, and that the proposition is true if Q belongs to this category and false if it belongs to the negation of this category (the complement of the set W).

Using this Axiom, the logical equivalence (7) can be transformed as follows:

$$\neg\left[\Diamond[x(Q)]\right] \equiv \Box\left[x(\text{Negation } Q)\right] \qquad (8)$$

If it is verified that

$$\neg \left[ \Diamond [x(Q)] \right] \equiv t \qquad (9)$$

then the equivalences (7) or (8) can be considered as partial mathematical models of Holmes' statement. In fact, this mathematical result seems too simple because equivalence (9), due to Łukasiewicz functional definitions of negation $\neg$ and possibility $\Diamond$, implies that $\Diamond[x(Q)] \equiv f$ and, hence, that $x(Q) = f$. Does this excessive simplicity disqualify the usefulness of the logical model? If one considers that the problem for Holmes is to develop a research process able to abolish uncertainty, the answer is negative because the statement $x(Q) = f$ represents the success of such process. In fact, the introduction of the third truth-value in Łukasiewicz logic was essential to guide the neural modeling of modal logic. As we are going to see in the next Section, modalities can be computed in the matrix neural models using two different ways to represent the uncertain truth-value in the network: on the one hand, defining a specific "truth-vector" to characterize uncertain truth-values (Section 4.2), and on the other, as a probabilistic weighting of the "true" and "false" vectors (Section 4.3).

The modal formats previously established are extremely important for the interpretation of Holmes' maxim, but they do not solve our problem because they do not constitute cognitive models and probabilities are absent. However, the modal relation (5) and the corresponding truth-valued representations (7) to (9) give us a powerful formal construction that shares with Holmes' postulate the enigmatic "feeling" of correctness that it produces. In fact, this formalism is a fundamental link to connect these modal equivalences with mathematical models of logical operations derived from neural models. Consequently, this approach covers some logical aspects of the statement, but a more comprehensive modeling should include the following points: a) a connection with the neural structures that produce and decode Holmes' maxim, and b) a link with probability concepts. As we are going to show in the next Section, matrix algebra is a fundamental framework for the construction of these models.

## 4. A neural approach to logic computations

The human brain is a spontaneous, pre-theoretical, computing device capable of performing sophisticated information processing, including mathematical and geometrical calculations. It is "pre-theoretical" in the sense that the human brains display many computational performances with no need of any explicitly programmed procedures or techniques. Clearly, language is used to express logical and mathematical computations whose cognitive bases still remain unexplained. We can ask ourselves if language acquisition involves a kind of implicit logical and mathematical programming that could explain such performances.

It is a common experience that the brain can solve problems concerning visual patterns without using any pre-existing mathematical knowledge. These problems are usually stated verbally. To illustrate this point, we can consider the image of Figure 1 representing a road.

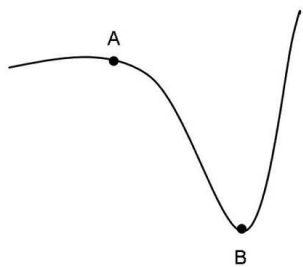

**Figure 1.** The points A and B over the design of a curved road have very different curvatures.

If you ask adults in which zone, A or B, the road shows larger curvature, the majority of the answers indicate point B. Of course, this conclusion is not based on the use of the classical mathematical formula

$$K(x) = \frac{d^2y}{dx^2}\left[1+\left(\frac{dy}{dx}\right)^2\right]^{-\frac{3}{2}}$$

that gives the curvature of point (x, y) in a plane trajectory with Cartesian equation y(x). On the contrary, almost surely this equation was strongly inspired by the pre-existing cognitive notion of curvature.

Logical judgments usually integrate the repertoire of human cognition, and even if many human actions are not submitted to the logic, the logical procedures are used for some crucial tasks that involve rational decisions. Consequently, it is particularly relevant to investigate the relation between cognitive models and logical performances (for an analysis of this point, see Binazzi 2012).

It is important to note that the three-valued logic, defined in Section 3, assumes the existence of well pre-classified data according to three cognitive categories: true, false, uncertain. The possibility of this classification is the consequence of clear diagnoses about the nature of the facts (e.g.: due to the lack of documents some historical facts can be transitorily classified in these three categories). But an open and evolving inquiry is full of transitory conjectures. In this situation, a sharp classification is not possible and the investigators must explore actual facts and build up their conclusions trying to decide if the conjectures are true or false in an environment full of uncertainties. In fact, this is the normal situation for many of the decisions adopted by humans in their natural environments.

### 4.1 Matrix Associative Memories

Many different approaches have been proposed to describe different aspects of neural function (Arbib 1995). In particular, the relations between brain dynamics and linguistic processes have been the subject of important investigations (see, for instance, Blutner 2004, beim Graben 2008). Let us mention the importance of matrices in the development of the theory of neural associative memories, a theory mainly developed around 1970 (Anderson 1972, Kohonen 1972, Cooper 1974). The matrix associative memories are able to model important facts about biological memories established in many experimental and clinical investigations. Comprehensive descriptions of this theory are included in Kohonen (1977) and Anderson (1995). These matrix memories are considered "distributed memories" because the residence of the information is a large set of synaptic contacts between neurons. This information is scattered and partitioned prior to be stored. This fact produces a desirable robustness of the model, in the sense that stored data can persist even in the presence of damages that produce a loss of neurons or synapses; this robustness of the model is a desirable fact because biology had revealed the existence of a relative tolerance to damage in real memories (Anderson 1995).

This theory assumes that the cognitive patterns correspond to neural activities that map on large dimensional vectors. A memory is a matrix that associates pairs of column vectors $(g_i, f_i)$, $i = 1, 2, ..., K$, where $f_i$ corresponds to the pattern i that enters the memory (e.g. the image of a person), and $g_i$ is the associated output (e.g., a name associated with the input image). As the theory shows, these vectors are composed by the electrochemical signals used by neurons to code information; these signals are generated in parallel by thousands of firing axons (Anderson 1995). The simplest form of the matrix that stores those pairs of vectors is as follows:

$$A = \sum_{i=1}^{K} g_i f_i^T \quad . \tag{10}$$

(the superindex T indicates transposition). Usually it is assumed that the set of stored input vectors is orthonormal (i.e. the $f_i$ are orthogonal between them and with lengths equal one). This assumption implies that the similarity between patterns is measured by the angle (equal patterns are parallel and completely different patterns are orthogonal). When a pattern $f_k$ enters the memory A, it is processed and generates an output. The following equation illustrates the mechanism:

$$A f_k = \sum_{i=1}^{K} \langle f_i, f_k \rangle g_i \qquad (11)$$

where $\langle f_i, f_k \rangle = f_i^T f_k$ is the scalar product (an operation that directly produces the cosines of the angle between this multidimensional unitary vectors; this cosines measure the angle and, consequently, the similarity between the patterns). If the input pattern belongs to the set stored into the memory, i.e. $f_k \in \{f_i\}$, we have

$$A f_k = g_k, \qquad (12)$$

a perfect association.

To include semantic contexts in the framework of this theory, different approaches involving a complex integration between inputs and contexts have been proposed (Arbib 1995). One of these approaches (Mizraji and Lin 2011) uses the Kronecker product to integrate inputs and contexts. In this framework, the matrix memory can be expressed as

$$M = \sum_{i,j} g_{ij} (f_i \otimes p_j)^T \qquad (13)$$

where $p_j$ is the context associated with the input $f_i$, and $g_{ij}$ is the output associated with the contextualized input. The symbol $\otimes$ represents the Kronecker (or tensor) product; in Appendix 2 we describe the basic properties of this operation (we use this tensor product inside a neural model, but for a foundation of this operation based on cognitive science see Smolensky 1990).

According to the algebraic rules involved in matrix algebra and Kronecker products (Graham 1981), the response of matrix memory M in the presence of an input and its context is

$$M(f_k \otimes p_h) = \sum_{i,j} \langle f_i, f_k \rangle \langle p_j, p_h \rangle g_{ij} \qquad (14)$$

with exact associations if the sets $\{f_i\}$ and $\{p_j\}$ are orthonormal, and if $f_k \in \{f_i\}$ and $p_h \in \{p_j\}$.

If matrix memories (10) and (13) represent biological associative memory modules, they are usually rectangular matrices of large dimensionality.

## 4.2 A Matrix-Vector Logic

The neural models previously described, provide us with a simple and powerful way to represent a large variety of logical operations[1]. Based on these memory modules, some years ago a matrix formalism named 'vector logic', that connects elementary propositional and modal logics with matrix neural models, was developed (see, for instance, Mizraji 2008). Inside this neural theory, the logical gates map on matrices and the truth values on vectors. The procedure to create the maps begins by mapping the truth values on orthogonal unitary vectors. Inside this neural formalism, the number of vector truth-values is *a priori* only limited by the dimensionality of the neural vectors. A large variety of many-valued logics can in principle be developed and sustained by matrix memory modules, because normally it is biologically plausible to assume that a neural vector has hundreds or thousands of components (Anderson 1995). In what follows, we describe two- and three-valued matrix-vector logics. In Section 4.3 we are going to show how a translation of the classical neural representation of modalities described by McCulloch-Pitts (1943) neuronal circuits can be adapted to produce an infinite-valued logic from two truth values and probabilistic weights.

---

[1] For an historical account of the links between logical theory and neural models see Eduardo Mizraji (2013) *En Busca de las Leyes del Pensamiento* (Second Edition), Montevideo: Trilce-Dirac.

Thus, a two-valued logic requires mapping $t \mapsto s$ and $f \mapsto n$, with s and n being orthonormal q-dimensional vectors; hence $\tau_2 = \{s, n\}$. Using this vector representation for the truth-values, the monadic and the dyadic two-valued gates respectively become the functions

$$\text{Mon}(2) : \tau_2 \to \tau_2 \quad,$$
$$\text{Dyad}(2) : \tau_2 \times \tau_2 \to \tau_2 \quad.$$

A three-valued logic is defined over the vector set $\tau_3 = \{s, n, h\}$, where h is a q-dimensional normal vector (orthogonal to vectors s and n) corresponding to the uncertain truth-value u ($u \mapsto h$). This vector set allows building up matrix versions for monadic and dyadic three-valued logical operators:

$$\text{Mon}(3) : \tau_3 \to \tau_3 \quad,$$
$$\text{Dyad}(3) : \tau_3 \times \tau_3 \to \tau_3 \quad.$$

As simple examples of Mon(2) we have the following matrices, $I_2$ and $N_2$, that correspond in this matrix framework to the logical two-valued Identity and Negation $\neg$:

$$I_2 = ss^T + nn^T, \tag{15}$$

$$N_2 = ns^T + sn^T. \tag{16}$$

Important examples of Dyad(2) operators are the matrix conjunction $C_2$ and disjunction $D_2$, given by the expressions:

$$C_2 = s(s \otimes s)^T + n(s \otimes n)^T + n(n \otimes s)^T + n(n \otimes n)^T, \tag{17}$$

$$D_2 = s(s \otimes s)^T + s(s \otimes n)^T + s(n \otimes s)^T + n(n \otimes n)^T. \tag{18}$$

Using these equations it can be easily proved that these operators execute vector versions of the classical operations. For instance,

$$C_2(s \otimes s) = s; \quad C_2(s \otimes n) = C_2(n \otimes s) = C_2(n \otimes n) = n.$$

It is important to see that the monadic operators (15) and (16) are particular cases of memory modules (10), and that the dyadic operators (17) and (18) correspond to memory modules (13). The vector logic provides explicit expressions for some of the Mon(3) and Dyad(3) matrix operators. We mention the fact that, under this formalism, the modalities possibility and necessity become very simple monadic matrices (a consequence of the Łukasiewicz functional definitions described in Section 3). Thus, for the three-valued vector system of logic, the identity $I_3$, the negation $N_3$, the matrix Pos (that represents in this formalism the possibility $\Diamond$) and the matrix Nec (that represents the necessity □) can be expressed by the following simple formulas:

$$I_3 = I_2 + hh^T, \tag{19}$$

$$N_3 = N_2 + hh^T, \tag{20}$$

$$Pos = I_2 + sh^T, \tag{21}$$

$$Nec = I_2 + nh^T, \tag{22}$$

where $I_2$ and $N_2$ are the operators given by equations (15) and (16)

Inside this matrix formalism, some basic theorems on logical modalities can be expressed as vector-matrix equalities. For instance, the postulate expressed by the equivalence (3) and its algebraic version given by (6), can be expressed by the matrix equation

$$Pos\ Val(Q) = N_3\ Nec[N_3\ Val(Q)]\ , \tag{23}$$

with Val(Q) representing the truth-value assigned to proposition Q, $Val(Q) \in \{s, n, h\}$. In addition, in this case the rules of matrix calculus allow us to express Aristotle's postulate as a product between the matrix operators involved:

$$Pos = N_3\ Nec\ N_3\ , \tag{24}$$

an identity not dependent on any particular value of the logical variable. These modal operators can be used as a way to represent logical modalities in terms of memory modules; for a discussion of the biological situations where these representations can be operative see Mizraji (2008).

The point of contact with Holmes' maxim is given by the matrix versions of equations (7) and (9):

$$N_3 \, Pos \, Val(Q) = Nec \, [N_3 \, Val(Q)] \qquad (25)$$

and

$$N_3 \, Pos \, Val(Q) = s \; . \qquad (26)$$

Consequently,

$$Nec \, [N_3 \, Val(Q)] = s \; . \qquad (27)$$

We can adapt the Axiom of Section 3 to this context by writing

$$N_3 \, Val(Q) = Val \, (Negation \, Q) \; . \qquad (28)$$

Note that Q is not a vector-matrix variable and we cannot apply the matrix negation on it. Instead, Negation Q is a proposition with a vector truth-valuation Val(Negation Q). Under this condition, equation (18) can be restated as

$$N_3 \, Pos \, Val(Q) = Nec \, [Val \, (Negation \, Q)] = s \; . \qquad (29)$$

This is a way to express that the impossibility of a proposition is equivalent to the necessity of the complement of that proposition. This result establishes a close contact with Holmes' maxim.

Obviously, equation (27) implies an extremely simple fact: Val (Negation Q) = s. As we mentioned previously, this seems a trivial conclusion, but it can be the result of a non-trivial process that was capable of eliminating uncertainty and of converging to this assertion. In fact, the valuation represented by Val(Q) (or Val(Negation Q)) involves a neural substrate apt to sustain the very complex cognitive process required to diagnose the truth-value of a proposition.

Equation (27) helps us to put the modal logical apparatus in terms of neural models, and to provide a new point of view to approach the spontaneous understanding that we feel when we are confronted with Holmes' old maxim. But this formalism does not consider any probabilistic evaluation of propositions, except the global assignment of uncertainties via the inclusion of a third truth-value. In the next Section, we are going to connect the previous matrix-logical approach with matrix models for associative memories capable of sustaining probabilistic operations.

It is easy to see that the matrix logical operations of this Section are particular cases of the Anderson-Kohonen distributed memories. In equation (25) Holmes' maxim is represented using modal logical operators and truth-evaluations expressed with a matrix-vector formalism. These logical operators are interpretable as memory modules that integrate a modular network (in fact a network of networks, because each one of the logical modules is by itself a neural network).

## 4.3 Guessing probabilities from "Neuro-Logic"

Many decisions are taken in the presence of uncertainty. These decisions usually rely heavily on modal operations. The decision adopted out of a group of choices must take into account a set of evaluations about possibilities. For instance, a gambler may say, 'I play roulette if I believe it is possible to win', and by gambling the risk-taker intentionally ignores the mathematical odds against winning. Considering this propensity, it is natural that Holmes' maxim was embedded in its modal frame.

However, Conan Doyle enlarges and refines this frame including a subjective probabilistic evaluation: *"whatever remains, however improbable"*. Probability is a well defined mathematical construct (see Feller 1968) and an interesting open problem is the accuracy of subjective estimations of a probability (for an important analysis of this point, see Pearl 2000). Nevertheless, we are continuously guessing probabilities, using a lot of accessible databases that help us to roughly estimate frequencies. Probability as a mathematical construct is one of the basis of the Bayesian explanation for Holmes' maxim sketched in Section 2, where the *a priori* probabilities involved are subjective guesses. How to deal with such probabilistic guesses in a logical theory? Many theoretical approaches connecting logic and probability have been published (e.g. Boole 1854, Keynes 1921) and these approaches have been connected with the problems of plausible reasoning (in particular by Polya 1990).

The neural models of logical operators have the potentiality to permit two different ways for the computation of the logical modalities. One way has been described in Section 4.2, with uncertainty "conceptualized" by a specific vector h. The other way is based on a recursive approach to logical modalities. We are going to adopt the vector-logic formalism to establish a connection with cognitive estimated probabilities and the recursive modal logic formalism. In the framework of this "neuro-logic", a way of introducing a probability guess in the logical formalism is to assign a numerical weight to the truth-value of an uncertain proposition. For instance, somebody can enunciate the proposition *"giving the clouds I am seeing, and the direction of the wind, I can forecast rain for the next two hours"* and based on my own experience about weather (i.e.: screening the databases installed in my memory modules), I can establish that such a proposition has an 80 % of probability of being true. This assignment (obviously not a probabilistic measure but a conjecture) in some way touches probability theory because it implies that the complementary situation (no rain) has a conjectural probability of 20 %. In the framework of the vector formalism that procedure can be modeled assuming that

$$\text{Val}(Q) = \alpha s + \beta n, \quad \alpha, \beta \in [0,1], \quad \alpha + \beta = 1, \tag{30}$$

s and n being the vector truth-values defined in Section 4.2.

Let us recall how inside the algebraic logic, recursive processes based on disjunction and conjunction (see Blanché 1968) can define the classical modalities. In this abstract approach, it is assumed the existence of an infinite set Q of propositions $Q_i$. This set can be mapped on a set of binary evaluations $\{q_1, q_2, ..., q_n, ...\}$, with $q_i = \text{Val}(Q_i) \in \{t, f\}$. The proposition "Q is possible", $\Diamond(Q)$, can be symbolically represented by

$$\Diamond(Q) = q_1 \vee q_2 \vee .... \vee q_n \vee ....,$$

that is an informal representation of the recursive process

$$\Diamond_{n+1}(Q) = q_{n+1} \vee \Diamond_n(Q) \quad n = 1, 2, ... \tag{31}$$

with $\Diamond_1(Q) = q_1$. In this process, the possibility $\Diamond(Q)$ is the limit of $\Diamond_n(Q)$ for $n \to \infty$. The symbol $\vee$ represents the dyadic disjunction.

In this formalism, the necessity is defined as follows. The proposition "Q is necessary", $\Box(Q)$, can be represented using a concatenated conjunction $\wedge$,

$$\Box(Q) = q_1 \wedge q_2 \wedge .... \wedge q_n \wedge ....,$$

or by the limit for $n \to \infty$ of the recursive process

$$\square_{n+1}(Q) = q_{n+1} \wedge \square_n(Q) \, , \, n = 1, 2, \ldots \tag{32}$$

with $\square_1(Q) = q_1$.

It is important to mention that these recursive processes reported in Mizraji (2008) using a matrix-vector formalism, were originally implemented by McCulloch and Pitts (1943) with formal neurons capable of executing OR and AND. Obviously, in the context of any neural model that pretends to describe a physical reality, recursions become finite. The formalism described in the previous section allows to represent these recursions using the conjunction and disjunction matrices (17) and (18). Let Nec (Q) describe a neural system that recursively evaluates possibilities exploring the information stored in a finite set $\{Q_i\}$ of propositions evaluated by vector truth-values $u_i = \text{Val}(Q_i)$. The matrix version of this process is as follows:

$$\text{Pos}_{n+1}[u] = D_2(u_{n+1} \otimes \text{Pos}_n[u]) \tag{33}$$

with $\text{Pos}_1[u] = u_1$, in general being $u_i = \alpha_i s + (1 - \alpha_i)n$, $\alpha_i \in [0,1]$. If we project this recursive process on vector s (the projection of a vector u on s is given by the scalar product $s^T u$) we obtain some interesting results. The scalar projection of $\text{Pos}[u] = \lim_{n \to \infty} \text{Pos}_n[u]$ is given by the product

$$s^T \text{Pos}[u] = 1 - (1 - \alpha_1)(1 - \alpha_2)(1 - \alpha_3) \ldots \, . \tag{34}$$

For a large number of data stored in a memory, this product can be approximated by a quasi-infinite recursion, and interpreted as a geometrical mean expression:

$$s^T \text{Pos}[u] \approx \lim_{n \to \infty} [1 - (1-\alpha)^n] . \tag{35}$$

Consequently

$$s^T \text{Pos}[u] \approx \begin{cases} 0 \text{ iff } \alpha = 0 \\ 1 \text{ iff } \alpha \neq 0 \end{cases} .$$

For the necessity operation, the matrix version is:

$$\text{Nec}_{n+1}[u] = C_2(u_{n+1} \otimes \text{Nec}_n[u]) \tag{36}$$

with $\text{Nec}_1[u] = u_1$ ($u_i = \alpha_i s + (1-\alpha_i)n$, $\alpha_i \in [0,1]$). Using the previous quasi-infinite approximation, the scalar projection of $\text{Nec}[u] = \lim_{n \to \infty} \text{Nec}_n[u]$ gives the product

$$s^T \text{Nec}[u] = \alpha_1 \alpha_2 \alpha_3 \ldots \tag{37}$$

Note that, due to the fact that u is the result of a recursive process, we have the following result:

$$s^T \text{Nec}[N_2 u] = (1-\alpha_1)(1-\alpha_2)(1-\alpha_3)\ldots = \beta_1 \beta_2 \beta_3 \ldots \tag{38}$$

This recursion can be averaged using the limit geometrical mean

$$s^T \text{Nec}[u] \approx \lim_{n \to \infty} \alpha^n . \tag{39}$$

that gives the classical scalar expression for the necessity

$$s^T \text{Nec}[u] \approx \begin{cases} 0 \text{ iff } \alpha \neq 1 \\ 1 \text{ iff } \alpha = 1 \end{cases}.$$

Note that these modal operators calculated from two valued operators, but with "probabilistic" truth-values given by equation (30), satisfy the theorem:

$$\text{Pos}[u] = N_2 \text{Nec}[N_2 u]. \tag{40}$$

A version of equation (8) is given by

$$N_2 \text{Pos}[u] = \text{Nec}[N_2 u] = s, \tag{41}$$

that gives us another formal "neuro-logical" version of Holmes' maxim, but now with the possibility of establishing contact with subjective probabilities (obviously only estimated probabilities from the mathematical point of view).

Remark that, in the scheme of this section, a research process implies the existence of a fact F that must be explained from a potential set of causes $Q = \{Q_i\}$. We can assume that *a priori* highly improbable causes do not belong to Q; consequently, they belong to the complement or negation of Q, that we represent symbolically by Negation Q. Remark that the elements of the set Q are not necessarily unlinked nor exhaustive: (a) they can be linked and (b) they are not exhaustive. Concerning (a), they can be linked because if, for instance, $Q_3$ represents the name of a possible guilty of a crime (say Jean) then $Q_7$ (say Jacques) can be the name of the same criminal (Jean-Jacques) or the name of his associate, both corresponding to the searched cause. Concerning (b), we mention that in general we may expect that $\sum_i \text{Prob}(Q_i) \neq 1$. The assumption here is that

$$\text{Val}(Q_i) = \alpha_i\, s + \beta_i\, n, \quad \alpha_i, \beta_i \in [0,1], \quad \alpha_i + \beta_i = 1. \tag{42}$$

Hence, $\text{Prob}(Q_i) = \alpha_i \leq 1$. This is the only assumption concerning probabilities. It implies a kind of conservation inside the judgment (conservation mapped in the complementarity of assigned probabilities for the two canonical truth-values s and n, and in turn supported by the complementarity of the associated conceptual sets (Mizraji and Lin 2011)). In this work we will leave as an open problem the link between these "cognitive probabilities" and the formal probabilities involved in the Bayesian treatment illustrated in Section 2.

Using equation (41) we can now establish a modal-probabilistic version of Holmes' maxim. Let us rewrite this equation as follows:

$$N_2\, \text{Pos}[e] = \text{Nec}[N_2\, e], \tag{43}$$

with e being a composed event described by a vector. This vector e emerges from the recursive process triggered by the search of evidences (evidences obtained from previous knowledge of the researcher or from fresh data coming from the external reality). We define the link between modalities and the corresponding binary probabilities as follows:

$$\text{Prob}(u) = s^T \text{Nec}[u]$$

For the modal evaluation describing Holmes' maxim, we have

$$N_2 \text{Pos}[e] = s \Rightarrow \text{Prob}(e) = 0$$
$$\text{Nec}[N_2 e] = s \Rightarrow \text{Prob}(N_2 e) = 1.$$

This is the final point of our neural version of Holmes' old maxim. According to equation (38), we have $s^T \text{Nec}[N_2 e] = (1-\alpha_1)(1-\alpha_2)(1-\alpha_3)... = \beta_1 \beta_2 \beta_3 ...$, then $\text{Prob}(N_2 e) = 1$ implies $\beta_i = 1, \forall i$. This very simple conclusion indicates that, even if the pre-judged probabilities $\beta_i^*$ were very small, the *a posteriori* result, after a research process showing the events $e_i$ to be impossible, induces the necessity of the complementary events and rise their probability to $\beta_i = 1$.

## 5. Discussion

From the very beginning of the mathematical theory of neuronal networks, the relation between logical reasoning and its neuronal bases has been a subject of primordial interest. The pioneering work by McCulloch and Pitts (1943) shows how some basic logical gates as NOT, AND and OR can be represented on the basis of binary neuronal elements by adjusting the thresholds and the synaptic weights of these formal neurons; other gates (e.g.: XOR) required a small network to be computed. In addition, McCulloch and Pitts showed that modalities ◊ and □ could be computed with recurrent networks based on the neuronal gates OR and AND respectively. These McCulloch-Pitts "logical neurons" had strong influence in the very important works published by Kleene (1951) and von Neumann (1956). Recently, the investigation of the links between reasoning and neural models acquired a new impulse and new perspectives promoted in part from the advances in cognitive sciences and the interest in neural computations. Between these new approaches, we want to mention the panoramic contribution of Stenning and van Lambalgen (2008, in particular Chapter 8) linking logic, cognition and biological evolution, and the investigations of d'Avila Garcez, Lamb and Gabbay concerning the representation by artificial neural networks of a variety of symbolic logical processes, including modalities and temporal logic (d'Avila Garcez 2007, d'Avila Garcez, Lamb and Gabbay 2009).

In the present work, we opted to use models based on Anderson-Kohonen matrix memories, where the logical computations are performed by networks of interconnected matrix modules (Mizraji and Lin 2011). In turn, each modular unit is composed by a large set of interconnected neurons represented by high dimensional matrices and can be programmed or instructed via a learning algorithm (eg: Widrow-Hoff algorithm, see Anderson 1995, chapter 9). This kind of matrix models present many aspects that enhance their biological plausibility (a point analyzed with detail in Anderson 1995), including their reliability in the presence of failures, their ability to create statistical averages from their learned inputs (interpretable as a source of conceptualizations, see Cooper 1974), and also their capacity to sustain logical gates and to display many-valued logics in the presence of uncertain data. It is now important to challenge these models with interesting problems linked with natural neural computations ("natural" in the sense that an adult human brain with a normal education and linguistic performances can do these computations). The Holmes' old maxim is one of these challenges and we think that these models provide interesting answers.

Using the neural-inspired logical formalism (Mizraji 2008 and Mizraji and Lin 2011), we showed in equation (29) how Holmes' maxim is represented using modal logical operators and how truth-evaluations are expressed with a matrix-vector formalism. As was previously mentioned, these logical operators are interpretable as memory modules that integrate a modular network (in fact a network of networks, because each one of the logical modules is by itself a neural network). The final point of our argument, equation (43), involves the matrices $N_2, Pos, Nec$ and the vector e. The matrix operations implement the abstract algebraic relationships constructed in order to formalize a basic modal logic, using the Boolean theoretical framework. Clearly, these operators are simplified models of biological neural devices. In this formalism, vectors are the 'stuff cognitive decisions are made on' inside the neural realm. Obviously, if mathematical approach helps us to explain our cognitive acceptability of Holmes' maxim, but on the other hand mathematics itself (being a cognitive construction) remains unexplained, then our eventual explanation can be considered incomplete and provisional, a flaw that we accept as a transitory step in our understanding of this kind of problem. We want to mention that many of the modern works concerning the neural representation of reasoning, emphasizes the need of nonmonotonic logics capable to deal with adaptive planning (see Stenning and van Lambalgen 2008). In Section 4.3 we saw how in our neural representation of Holmes' maxim, all the process –even if it is guided by the framework of modal logic truth-functionality – is dependent on the adaptive evaluations of probabilities. We can ask if, formally, this process represents a class of nonmonotonic reasoning (as described for instance in d'Avila Garcez, Lamb and Gabbay 2009, Chapter 2) but we do not have still a clear answer. Surely the possibility to formalize this process as a nonmonotonic logic is an issue that deserves further exploration.

The argument developed in this work suggests that we accept Holmes' maxim as true because our brains are capable to activate neural modules able to perform modal logical computations. Our formal approach is neutral in the debate about the "Nature *versus* Nurture" origin of our logical abilities: these neural logical modules can be the result of a genetically coded ontogenetic process, or, on the contrary, they can be the result of a learning process occurring in a particular cultural environment (for a discussion of this point, see Mizraji and Lin 2011). We argue that the spontaneous computation involved in the understanding of Holmes' maxim is only one example among many others; all of them emerging from a natural biological design that obviously includes our cognitive brains, together with our sensory and motor systems. In fact, language uses some computational codes that trigger complex cognitive procedures. For instance, a preposition like "in" can induce the brain to represent a complex spatial relationship between an object and a container. Prepositions and logical words are crucial linguistic constructions that can act as passwords that give access to sophisticated neurocomputational operations.

APPENDIX 1.

**Sherlock Holmes' old maxim.**

We transcribe three well-known versions of Holmes' maxim (Doyle, Penguin Edition 1981). In "The Adventure of the Beryl Coronet", Holmes says: *"It is an old maxim of mine that when you have excluded the impossible, whatever remains, however improbable, must be the truth"*.

Another version is in the novel "The Sign of Four" where Holmes says to Dr. Watson: *"How often I said to you that when you have eliminated the impossible, whatever remains, however improbable, must be the truth?"*.

The last version we want to reproduce here is the one included in "The Adventure of the Bruce-Partington Plans", where Holmes comments: *"We must fall back upon the old maxim that when all other contingencies fail, whatever remains, however improbable, must be the truth"*.

APPENDIX 2

**Kronecker Product** (Graham 1981).

For the matrices $A = [a_{ij}] \in R^{m \times n}$ and $B = [b_{ij}] \in R^{p \times q}$, the Kronecker product is a matrix $A \otimes B \in R^{mp \times nq}$ defined by

$$A \otimes B = [a_{ij} B].$$

Some important properties of the Kronecker product are the following:

A1. $(A + A') \otimes (B + B') = A \otimes B + A \otimes B' + A' \otimes B + A' \otimes B'$

A2. $(A \otimes B)^T = A^T \otimes B^T$

A3. $(A \otimes B)(A' \otimes B') = (AA') \otimes (BB')$

**Acknowledgments**: I thank Julio A. Hernández, Juan C. Valle-Lisboa and Andrés Pomi for discussions and comments. This work was partially supported by PEDECIBA (Uruguay), CSIC, (UdelaR).